\documentclass[aps,prb,floatfix,twocolumn,showpacs]{revtex4} 

\usepackage{epsfig,amsmath,amssymb}

\begin{document}

\title{Theory of Single Electron Spin Relaxation in Si/SiGe Lateral Coupled Quantum Dots}
\author{Martin Raith$^1$, Peter Stano$^{2,3}$ and Jaroslav Fabian$^1$}
\affiliation{$^1$Institute for Theoretical Physics, University of Regensburg, D-93040 Regensburg, Germany\\
$^2$Physics Department, University of Arizona, Tucson, Arizona 85716, USA\\
$^3$Institute of Physics, Slovak Academy of Sciences, 845 11 Bratislava, Slovakia}

\vskip1.5truecm

\begin{abstract}
We investigate the spin relaxation induced by acoustic phonons in the presence of spin-orbit interactions in single electron Si/SiGe lateral coupled quantum dots. The relaxation rates are computed numerically in single and double quantum dots, in in-plane and perpendicular magnetic fields. The deformation potential of acoustic phonons is taken into account for both transverse and longitudinal polarizations and their contributions to the total relaxation rate are discussed with respect to the dilatation and shear potential constants. We find that in single dots the spin relaxation rate scales approximately with the seventh power of the magnetic field, in line with a recent experiment. In double dots the relaxation rate is much more sensitive to the dot spectrum structure, as it is often dominated by a spin hot spot. The anisotropy of the spin-orbit interactions gives rise to easy passages, special directions of the magnetic field for which the relaxation is strongly suppressed. Quantitatively, the spin relaxation rates in Si are typically 2 orders of magnitude smaller than in GaAs due to the absence of the piezoelectric phonon potential and generally weaker spin-orbit interactions.
\end{abstract}


\maketitle

\section{Introduction}
For more than a decade semiconductor quantum dots have been in research focus for quantum information processing.\cite{PhysRevA.57.120,PhysRevA.61.062301,Burkard:2007p2938,PhysRevB.72.201304,epub1823,epub7807} The experimental control over the electron spin in quantum dots has seen enormous progress, with lateral gated GaAs structures demonstrating the state of the art.\cite{RevModPhys.79.1217} Magnetic\cite{Nature-442-766} and electric\cite{K.C.Nowack11302007} coherent spin manipulations have been demonstrated, while the electron spin measurement has been achieved using spin to charge conversion techniques.\cite{PhysRevB.74.195305} For the latter, double dots\cite{pfund:252106} have proven especially useful exploiting the spin Coulomb blockade.\cite{PhysRevB.76.035315} However, the GaAs spin qubit seems to have reached its fundamental limit for the coherence which is due to the nuclear spins inherent in all III-V semiconductors.\cite{Nature-435-925,PhysRevLett.100.236802,TAP.115.183}

Materials composed of atoms without nuclear magnetic moment, such as Si and C, seem a natural solution for the problem of the nuclear induced decoherence.\cite{PhysRevLett.92.076401, NatureMat-8-383, PhysRevB.71.014401,RRL-3-61} That is why Si based quantum dots have recently seen a revived interest. Although the quantum dot technology is not yet as mature as in GaAs, several perspective setups are being actively pursued.\cite{Nature-Physics-4-540, zwanenburg:124314, lim:173502, PhysRevB.80.115331, PhysRevB.77.073310, RRL-2-59, 1367-2630-12-11-113019} We note that a spin to charge conversion was reported recently.\cite{PhysRevLett.106.156804} In addition to the absence of nuclear spins, Si seems potentially advantageous because of weaker spin-orbit interactions, promising less decoherence, and a stronger $g$ factor, allowing spin control at smaller magnetic fields.  

On the other hand, the electron effective mass in Si is larger than in GaAs, so Si dots must be smaller at a given orbital energy scale. In addition, and perhaps more seriously, a major issue for silicon based quantum computation is the valley degeneracy of its conduction band electrons.\cite{PhysRevB.82.155312,PhysRevB.81.085313} In the bulk, the conduction band minima are located at the $X$ valleys, that is at $k_v\approx 0.84 k_0$, $v=1,\ldots6$, toward the six $X$ points of the Brillouin zone, where $k_0 = 2\pi/a_0$ and $a_0=5.4\,\text{\AA}$ is the lattice constant.\cite{0268-1242-12-12-001} In a (001)-grown Si heterostructure the valley degeneracy is partially lifted due to the presence of the interface and/or due to strain,\cite{RevModPhys.54.437} leaving a twofold conduction band minimum, the $\pm z$ valleys, which are separated from the fourfold excited valley states by at least $10\,\text{meV}$,\cite{0268-1242-12-12-001,0268-1242-19-10-R02,PhysRevB.81.085313} large enough to neglect the upper four valleys.\cite{PhysRevB.82.155312,PhysRevB.81.085313,PhysRevB.81.115324} The remaining twofold valley degeneracy is lifted if the perpendicular confinement is asymmetric. Then the orbital wave functions become symmetric and antisymmetric combinations of the single valley states,\cite{PhysRevB.75.115318} which are separated by the energy difference called the ground-state gap\cite{PhysRevB.81.115324} (or valley splitting). 

In recent years the origin and possible control of the valley splitting has been in focus. Measurements in silicon heterostructures reveal a valley splitting of the order of $\mu\text{eV}$.\cite{Weitz1996542,0268-1242-12-4-007,Schumacher1998260,PhysRevB.67.113305,PhysRevLett.93.156805,PhysRevB.72.165429,RevModPhys.54.437} On the other hand, theoretical estimates of perfectly flat structures propose a splitting about three orders of magnitude larger.\cite{boykin:115} Taking into account detailed properties of the interface (e.g., roughness), experiment and theory come to an agreement,\cite{nphys475,friesen:202106,PhysRevB.19.3089,PhysRevB.74.245302,2006cond.mat..6395V,PhysRevB.75.115318,PhysRevB.81.115324,PhysRevB.80.081305,1367-2630-12-3-033039} and additional (in-plane) confinement allows the valley splitting to reach values of the order of $\text{meV}$.\cite{nphys475} In Si/SiO$_2$ systems, the splitting can even be tens of $\text{meV}$.\cite{Ouisse1998731,PhysRevLett.96.236801,PhysRevB.69.161304} Here we assume that the splitting is at least $1\,\text{meV}$ and we can use the effective single valley approximation,\cite{PhysRevB.81.085313,PhysRevB.80.205302} in which only the lowest valley eigenstate is considered. This choice is strengthened by the fact that electron spins in valley-degenerate dots would not be viable qubits.\cite{PhysRevB.82.155312,PhysRevB.81.085313,PhysRevB.80.205302}

In the single valley approximation, the Si dot resembles the fairly well understood GaAs one. The main goal of this article is to carry out a comparison on a quantitative level, providing realistic values for the electron spin relaxation as available for GaAs dots.\cite{PhysRevLett.100.046803, PhysRevB.64.125316, PhysRevB.75.195342, PhysRevB.74.045320,Nature-430-431}

The relevant sources of spin relaxation in GaAs quantum dots in a magnetic field are the electron-phonon couplings modeled by the piezoelectric and deformation potential theory.\cite{PhysRevLett.96.186602, PhysRevB.74.045320} Since silicon is not piezoelectric, only the deformation potential mechanism remains. From this point of view, a Si dot is closer to InGaAs than to GaAs, as in InGaAs the relative importance of the deformation versus the piezoelectric potential is enhanced due to a larger $g$ factor.\cite{sherman2005:PRB} However, differently from (In)GaAs, in which transverse acoustic phonons do not contribute to the deformation potential coupling, both transverse and longitudinal acoustic phonons cause spin relaxation in Si.\cite{PhysRev.101.944, PhysRevB.77.115438, PhysRevB.81.235326, pop:4998, 0957-4484-10-2-307} In the present work special attention is given to the importance of the transverse phonons and to the role of the dilatation and shear potential constants, $\Xi_{\text{d}}$ and $\Xi_{\text{u}}$ respectively, which parameterize the electron-phonon coupling strengths.

In Si, the spin relaxation/decoherence rates were computed perturbatively for single dot single electron\cite{PhysRevB.66.035314} and single\cite{2008PhRvB..77k5438P} and double\cite{culcer:073102} dot singlet-triplet transitions. Also the spin relaxation due to the
modulation of electron $g$ factor by the phonon-induced strain was investigated.\cite{PhysRevB.68.045308} Experimentally the rates were measured on quantum dot ensembles\cite{pan:013103,Tyryshkin2006257} and on a many-electron quantum dot,\cite{2009arXiv0908.0173H,PhysRevLett.104.096801} and a few electron quantum dot.\cite{PhysRevLett.106.156804} We obtain the spin relaxation rates non-perturbatively, using exact numerical diagonalization, for a wide range of magnetic fields and interdot coupling. Our results allow us to discuss regimes beyond the validity of perturbative treatments and to specify the accuracy of common approximations.

The article is organized as follows. In Sec.~II we define our model of the double quantum dot, the electron-phonon interaction, and the spin relaxation. In Sec.~III we briefly review the Si single and double dot spectra, paying attention to the states' orbital symmetries. In Sec.~IV.A we investigate the spin relaxation in a single dot in an in-plane magnetic field. Comparing with a recent experiment, we find that (i) the results of the experiment indicate that the main spin relaxation channel is the mechanism we study here, and (ii) the spin-orbit coupling strength of $\sim 0.1 \text{meV\AA}$ seems realistic for the Si/SiGe lateral quantum dots. We also present analytical formulas for the spin relaxation rate in the lowest order of the spin-orbit interactions. Comparing with exact numerics, we demonstrate that these formulas are quantitatively reliable up to modest magnetic fields of 1-2 T. We find that a further analytical simplification often adopted, the isotropic averaging of the interaction strengths, leads to a result correct only within an order of magnitude. In Sec.~IV.B we deal with the double dot case and demonstrate that here the spin relaxation rate is sensitive to the spectral anticrossings (spin hot spots),\cite{PhysRevLett.81.5624} especially if the magnetic field is perpendicular to the heterostructure. For in-plane fields, the anisotropy of the spin-orbit interactions leads to the appearance of easy passages---magnetic field directions in which the relaxation rate is quenched by many orders of magnitude.\cite{PhysRevB.74.045320,PhysRevLett.96.186602} These results are analogous to the GaAs quantum dots. Finally, we conclude in Sec.~V.

\begin{figure}
\centerline{\psfig{file=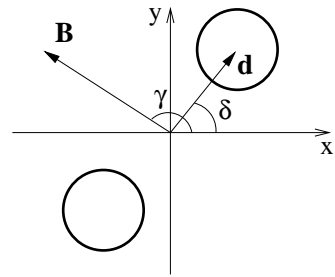,width=0.5\linewidth}}
\caption{Orientation of the double dot in the coordinate system $\hat x = \left[100\right]$, $\hat y = \left[010\right]$. The potential minima, sketched by two circles, are parameterized by the position vectors $\pm\mathbf{d}$ or by the distance $2d$ and the angle $\delta$. The in-plane magnetic field orientation is given by the angle $\gamma$.}
\label{Model:fig:dot_orientation}
\end{figure}

\section{Model}\label{Sect:Model}
Within the two-dimensional, effective mass and effective single valley approximations a laterally coupled, top-gated double quantum dot (DQD) in a silicon heterostructure with the growth direction $\hat z = \left[001\right]$ is described by the Hamiltonian
\begin{equation}\label{Model:Hamiltonian}
H = \dfrac{ \hbar^2 \mathbf{K}^2}{2m} + V(x,y) + \dfrac{g}{2} \mu_{\text{B}} \boldsymbol{\sigma} \!\cdot\! \mathbf{B} + H_{\text{br}} + H_{\text{d}},
\end{equation}
with $m$ the effective electron mass and $\mathbf{r}= \left(x,y\right)$ is the in-plane position vector. The vector of the kinetic momentum $\hbar \mathbf{K} = -\text{i} \hbar \nabla + e \mathbf{A}$, with $e$ the proton charge, consists of the canonical momentum and the vector potential in symmetric gauge, $\mathbf{A} = B_z \left(-y/2,x/2\right)$. We neglect the orbital effects of the in-plane magnetic field component, a good approximation up to roughly $10\,\text{T}$ for the usual heterostructures.\cite{epub8285} The in-plane coordinates are chosen to be the crystallographic axes, that is, $\hat x = \left[100\right]$ and $\hat y = \left[010\right]$. We use the biquadratic model\cite{PhysRevB.81.085313,PhysRevB.77.045310} for the electrostatic confinement potential,
\begin{equation}\label{Model:Potential}
V(x,y) = \frac{1}{2} m \omega_{0}^2 \, \text{min}\{\left(\mathbf{r}-\mathbf{d}\right)^2 \!,\left(\mathbf{r}+\mathbf{d}\right)^2\},
\end{equation}
where $\pm\mathbf{d}$ denote the positions of two potential minima. Alternatively, the minima are parameterized by the interdot distance $2d$ and the angle between the main dot axis $\mathbf{d}$ and $\hat x$, denoted by $\delta$. Characteristic scales are given by the confinement energy $E_0 = \hbar \omega_{0}$ and the confinement length $l_{0} = \left( \hbar / m \omega_{0} \right)^{1/2}$. The Zeeman energy in Eq.~\eqref{Model:Hamiltonian} is proportional to the effective Land\'{e} factor $g$ and the Bohr magneton $\mu_{\text{B}}$, while $\boldsymbol{\sigma} = \left( \sigma_x, \sigma_y, \sigma_z \right)$ is the vector of the Pauli matrices. The magnetic field is given by $\mathbf{B} = \left( B_{\parallel} \cos \gamma, B_{\parallel} \sin \gamma, B_z \right)$, where $\gamma$ is the angle between the in-plane component of $\mathbf{B}$ and $\hat x$. The geometry is plotted in Fig.~\ref{Model:fig:dot_orientation}.

Extrinsic spin-orbit coupling (SOC) leads to additional terms in the Hamiltonian of two-dimensional systems without inversion symmetry.\cite{epub1823} Structure inversion asymmetry arising, for example, due to an electric field applied along the growth direction, results in the Bychkov-Rashba Hamiltonian
\begin{equation}\label{Model:HBR}
H_{\text{br}} = \alpha \left(\sigma_x K_y - \sigma_y K_x \right),
\end{equation}
with an electrically tunable coupling parameter $\alpha$. The bulk inversion asymmetry of zinc blende semiconductors such as GaAs, is not present in silicon with diamond structure and the corresponding linear and cubic Dresselhaus interactions are absent. However, a silicon heterostructure is of $\text{C}_{\text{2v}}$ symmetry, so there is also a generalized Dresselhaus term,\cite{PhysRevB.69.115333,PhysRevB.77.155328}
\begin{equation}\label{Model:HDR}
H_{\text{d}} = \beta \left(-\sigma_x K_x + \sigma_y K_y \right) ,
\end{equation}
which is identical to the Dresselhaus Hamiltonian stemming from the bulk inversion asymmetry in III-V semiconductors. The parameter $\beta$ depends on the interface (step height) and well width (number of Si-atomic layers). An alternative parameterization utilizes the spin-orbit coupling lengths $l_{\text{br}}=\hbar^2/2m\alpha$ and $l_{\text{d}}=\hbar^2/2m\beta$. In this work we assume an asymmetric quantum well along $\hat z$ with $\alpha$ and $\beta$ of comparable strengths.\cite{PhysRevB.77.155328}

Our parameters for the numerical calculations are as follows: The system of interest is a (001)-grown SiGe/Si/SiGe quantum well, where the thin Si layer is sandwiched by the relaxed SiGe with a Germanium concentration of about 25\%. The bulk electron effective mass of the $X$ valleys is anisotropic in the direction transverse and longitudinal to the corresponding $k_v$-vector, given by $m_l=0.916m_e$ and $m_t=0.191m_e$, respectively, where $m_e$ is the free electron mass. The in-plane mass of the $z$ valley states is therefore the transverse mass. Due to the tensile strain (in-plane) of the Si layer the effective mass is slightly increased\cite{PhysRevB.48.14276} compared to the unstrained bulk Si and we use $m=0.198 m_e$.\cite{PhysRevB.48.14276} The effective Land\'e factor is $g=2$\cite{2009arXiv0908.0173H,nphys475} and the SOC strengths are set to be $\alpha=0.05 \,\text{meV\AA}$ and $\beta=0.15 \,\text{meV\AA}$, respectively.\cite{PhysRevB.77.155328} Our choice is based on results of theoretic tight-binding calculations of Ref.~\onlinecite{PhysRevB.77.155328}, as experimentally the SOC in silicon dots has not been measured up to date. We use the confinement energy $E_0 = 1 \,\text{meV}$ equivalent to the confinement length $l_{0} = 20 \,\text{nm}$, which corresponds to realistic dot sizes.\cite{2009arXiv0908.0173H,gandolfo:063710}

We consider transitions mediated by phonons, with the accompanying spin-flip allowed due to the presence of the spin-orbit interactions. In a (001)-grown quantum well, the electron-phonon coupling for intravalley scattering is the deformation potential of transverse acoustic (TA) and longitudinal acoustic (LA) phonons given by\cite{PhysRev.101.944,PhysRev.118.1523,PhysRevB.77.115438,PhysRevB.81.235326,pop:4998,0957-4484-10-2-307}
\begin{equation}\label{Model:Helph}
H_{\text{def}} = \text{i} \sum_{\mathbf{Q}, \lambda} \sqrt{\dfrac{\hbar Q}{2 \rho V c_{\lambda}}} D_{\mathbf{Q}}^{\lambda} \left[ b^{\dagger}_{\mathbf{Q},\lambda} \text{e}^{\text{i} \mathbf{Q} \cdot \mathbf{R}} \!-\! b_{\mathbf{Q},\lambda} \text{e}^{-\text{i} \mathbf{Q} \cdot \mathbf{R}} \right] \! ,
\end{equation}
with
\begin{equation}\label{Model:D}
D_{\mathbf{Q}}^{\lambda} = \Xi_{\text{d}} \mathbf{\hat e}_{\mathbf{Q}}^{\lambda} \cdot \mathbf{\hat Q} + \Xi_{\text{u}} \mathrm{\hat e}_{\mathbf{Q},z}^{\lambda} \hat Q_z ,
\end{equation}
where $\mathbf{Q} = \left( \mathbf{q}, Q_z \right)$ is the phonon wave vector, $\mathbf{\hat Q}$ is its unit vector and $\mathbf{R} = \left( \mathbf{r}, z \right)$ is the electron position vector. The summation includes all polarizations\cite{grodecka05a} $\lambda = \text{TA1, TA2, LA}$, and $c_{\lambda}$ is the corresponding sound velocity. The phonon annihilation (creation) operator is denoted by $b$ ($b^{\dagger}$) and the polarization unit vector reads $\mathbf{\hat e}$. The mass density is given by $\rho$ and $V$ is the volume of the crystal. The deformation potential strength is set by the dilatation and shear potential constants $\Xi_{\text{d}}$ and $\Xi_{\text{u}}$, respectively. Note that unlike in GaAs there is no piezoelectric phonon potential in Si.

We define the single-electron spin relaxation rate, which is the inverse of the spin lifetime $T_1$, as the sum of the transition rates from the upper Zeeman split ground state, called $\Gamma_{\text{S}}^{\uparrow}$ in the following, to all lower-lying states $\Psi_{\downarrow}$ with opposite spin. Each individual transition rate is evaluated using Fermi's Golden Rule in the zero-temperature limit\cite{PhysRevB.74.045320,epub7807,PhysRevB.77.045310}
\begin{equation}\label{Model:rate}
\Gamma_{\text{spin}} = \dfrac{\pi}{\hbar \rho V} \sum_{\mathbf{Q}, \lambda} \dfrac{Q}{c_{\lambda}} \left| D_{\mathbf{Q}}^{\lambda} \right|^2 \left| M_{\uparrow \downarrow} \right|^2 \delta(\omega_{\uparrow \downarrow} - \omega_{\mathbf{Q}}) ,
\end{equation}
where $M_{\uparrow \downarrow} = \langle \Gamma_{\text{S}}^{\uparrow} | \text{e}^{\text{i} \mathbf{Q} \cdot \mathbf{R}} | \Psi_{\downarrow} \rangle$ is the matrix element for the corresponding initial and final states and $\hbar \omega_{\uparrow \downarrow}$ is the energy difference between these states. We evaluate Eq.~\eqref{Model:rate} numerically using the parameters $\rho = 2.3 \times 10^3 \,\text{kg/m}^3$, and $c_{t} = 5 \times 10^3 \,\text{m/s}$ for TA, and $c_{l} = 9.15 \times 10^3 \,\text{m/s}$ for LA phonons. The choice of deformation potential constants is not unique.\cite{fischetti:2234} We use $\Xi_{\text{d}}= 5 \,\text{eV}$ and $\Xi_{\text{u}}= 9 \,\text{eV}$ according to Ref.~\onlinecite{LBGroupIII}, noting that other combinations such as $\left(\Xi_{\text{d}},\Xi_{\text{u}}\right)=$ $(1.1,6.8)\,\text{eV}$,\cite{pop:4998} $(1.13,9.16)\,\text{eV}$, and $(-11.7,9)\,\text{eV}$\cite{0957-4484-10-2-307} appear alternatively. The needed electron wave functions and energies are obtained numerically as eigensystem of the Hamiltonian in Eq.~\eqref{Model:Hamiltonian}, which we diagonalize with the method of finite differences using Dirichlet boundary conditions. The magnetic field is included by the Peierls phase and the diagonalization is carried out by the Lanczos algorithm. Typically we use a grid of 50$\times$50 points, which results in a relative precision in energy of $10^{-5}$ in zero magnetic field.

\section{Single Electron States}
\subsection{Energy spectrum in zero magnetic field}\label{Sect:ElSt_Bzero}
\begin{figure}
\centerline{\psfig{file=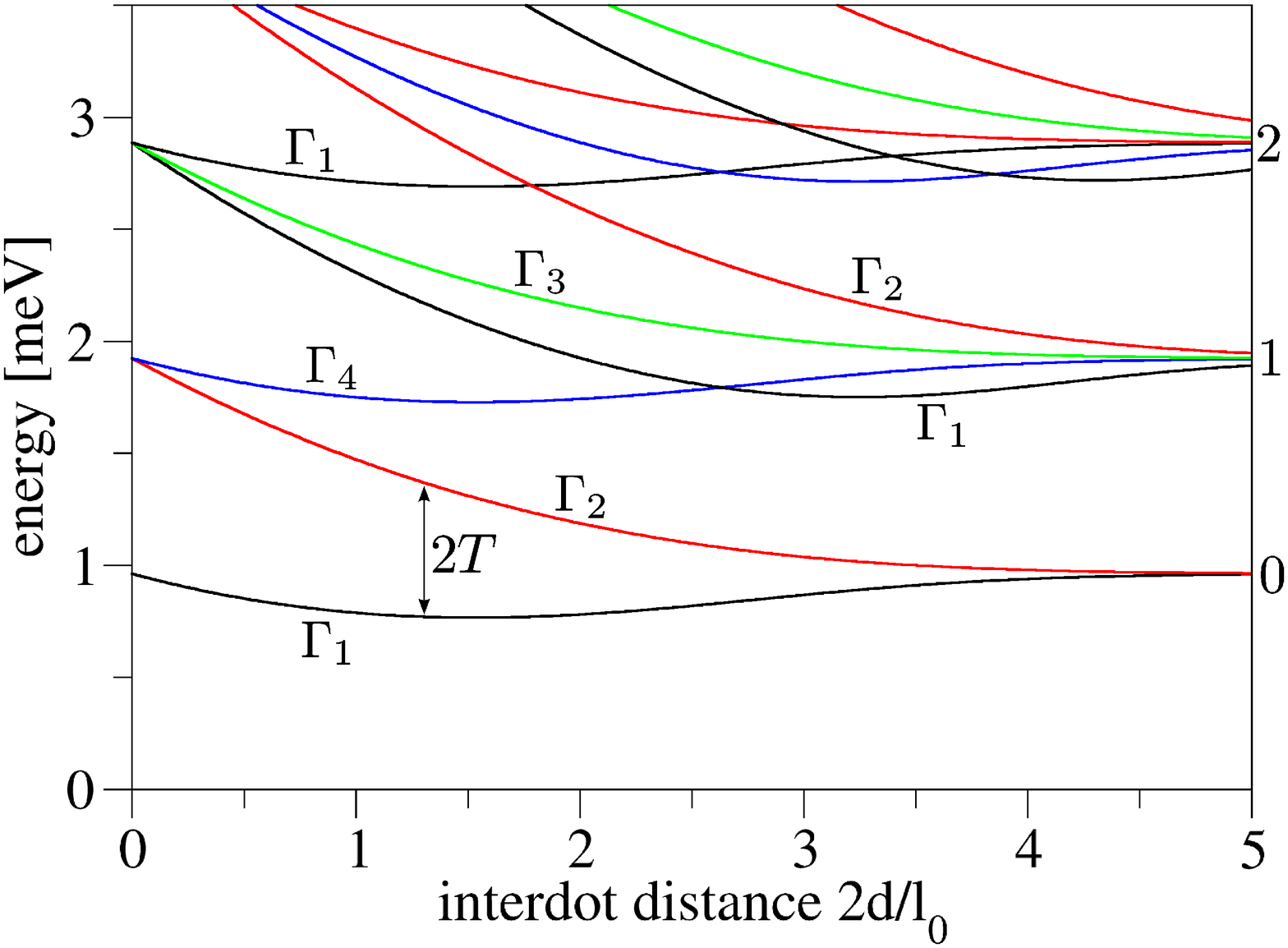,width=0.95\linewidth}}
\caption{(color online) Calculated energy spectrum of the Si double quantum dot with respect to the interdot distance at zero magnetic field. The states are labeled (colored) according to the irreducible representations $\Gamma_i$ of $\text{C}_{\text{2v}}$. On the right-hand side we give the highest orbital momentum of associated single dot states (Fock-Darwin states). The tunneling energy $T$ is also shown.}
\label{ElSt:fig:DQDspectrum}
\end{figure}

In order to understand the details of the spin relaxation in silicon DQDs, we review briefly their electronic properties in zero magnetic field, including group theoretical classification, the influence of SOC and the most important quantities for experiments. The Hamiltonian Eq.~\eqref{Model:Hamiltonian} for $\mathbf{B}=0$ and without SOC has $\text{C}_{\text{2v}}\otimes \text{SU(2)}$ symmetry. We can then label the orbital states according to the irreducible representations $\Gamma_i$, $i=1,...,4$ of the Abelian point group $\text{C}_{\text{2v}}$ noting that each state is doubly spin-degenerate due to $\text{SU(2)}$. This is done in Fig.~\ref{ElSt:fig:DQDspectrum}, where the energy spectrum vs.~the interdot distance in units of $l_{0}$ is plotted. Note that the potential $V$, Eq.~\eqref{Model:Potential}, was chosen such that the states converge to Fock-Darwin states\cite{Chakraborty1999} in the limit of zero or infinite interdot distance. In the following we focus on the intermediate region where the interdot distance is comparable to the confining length. This is typically the region of experimental interest, as well as the one in which numerics becomes indispensable. Here we find several level crossings which may be lifted in the presence of SOC. Such anticrossings, also called spin hot spots,\cite{PhysRevLett.81.5624} are of great importance for spin relaxation as we will see later. However, the linear SOC terms [Eq.~\eqref{Model:HBR} and Eq.~\eqref{Model:HDR}] do not lead to level repulsion in the first order although allowed by symmetry.\cite{PhysRevB.72.155410} We conclude that in zero magnetic field the DQD spectrum of silicon does not exhibit relevant spin hot spots.

For many applications including quantum dot spin qubits, the important physics happens at the bottom of the spectrum. We denote the spin-degenerate ground state as $\Gamma_1 \equiv \Gamma_{\text{S}}$ and the first excited state as $\Gamma_2 \equiv \Gamma_{\text{A}}$ to indicate the symmetry under inversion $I_{xy}$. The energy difference between these states is parameterized by the tunneling energy,\cite{PhysRevB.72.155410} $T=\left(E_{\text{A}} - E_{\text{S}}\right)/2$, a characteristic quantity for DQDs directly measurable experimentally.\cite{NanoLett-9-3234} Note that within the single valley approximation we assumed a valley splitting of at least $1\,\text{meV}$ which exceeds $2T$ at all interdot distances.

\begin{figure}
\centerline{\psfig{file=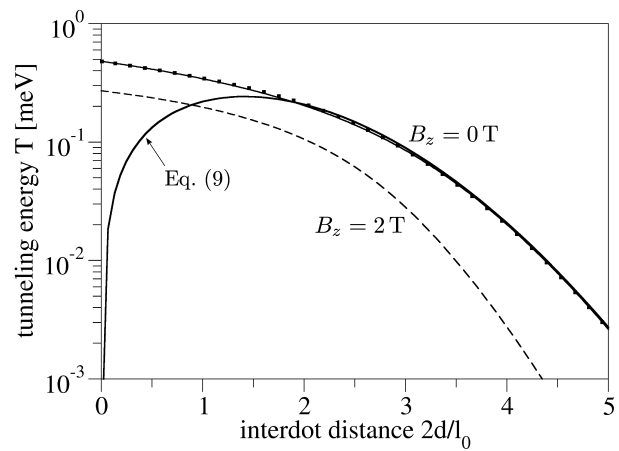,width=0.95\linewidth}}
\caption{Calculated tunneling energy vs.~interdot distance for zero magnetic field calculated by exact numerical diagonalization (dotted line), exact LCSDO formulas (thin solid line) and leading order approximation (Eq.~\eqref{ElSt:LCSDO-T-approx}, thick solid line). The tunneling energy for a finite perpendicular magnetic field ($B_z = 2\,\text{T}$, dashed line) is given for comparison.}
\label{ElSt:fig:tunnelingenergy}
\end{figure}

Using a linear combination of single dot orbitals\cite{PhysRevB.72.155410} (LCSDO) we can approximate the exact wave functions by analytical expressions. Let $\Psi_{n,l}(\mathbf{r})$ be the Fock-Darwin state (omitting spin), where $n$ is the principal and $l$ the orbital quantum number.\cite{Chakraborty1999} Then the lowest orbital eigenstates of the DQD can be approximated using the Fock-Darwin states centered at the potential minima as
\begin{eqnarray}\label{ElSt:LCSDO}
\Gamma_{\text{S}} &=& N_{+} \left[ \Psi_{0,0}(\mathbf{r}+\mathbf{d}) + \Psi_{0,0}(\mathbf{r}-\mathbf{d}) \right] , \nonumber \\
\Gamma_{\text{A}} &=& N_{-} \left[ \Psi_{0,0}(\mathbf{r}+\mathbf{d}) - \Psi_{0,0}(\mathbf{r}-\mathbf{d}) \right] .
\end{eqnarray}
Here $N_{\pm}$ are normalization constants. Calculating the eigenenergies as the expectation values of the Hamiltonian, Eq.~\eqref{Model:Hamiltonian}, for zero magnetic field and without SOC, we obtain the tunneling energy as plotted in Fig.~\ref{ElSt:fig:tunnelingenergy}. It is in excellent agreement with the exact numerical result. In the limit of large interdot distances the leading order reads
\begin{equation}\label{ElSt:LCSDO-T-approx}
T \approx E_0 \dfrac{d}{\sqrt{\pi}\,l_{0}} \text{e}^{-\left(d/l_{0}\right)^2},
\end{equation}
which is a good approximation if $2d/l_{0}>2.5$.

In principle SOC terms affect the tunneling energy. However, it was shown\cite{PhysRevB.72.155410} that this correction is of fourth order in the spin-orbit strengths $\alpha$ and/or $\beta$. For our parameters here it is of the order of $\text{peV}$ and therefore negligible for all experimental purposes.

\subsection{Energy spectrum in non-zero magnetic field}
\begin{figure}
\centerline{\psfig{file=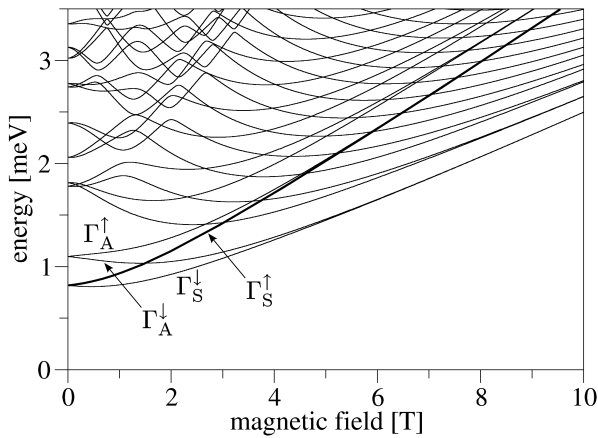,width=0.95\linewidth}}
\caption{Calculated energy spectrum of the DQD with interdot distance $2d/l_{0}=2.5$ plotted against the perpendicular magnetic field. The thick line indicates $\Gamma_{\text{S}}^{\uparrow}$, the lowest state with opposite spin-polarization as the ground state.}
\label{ElStwB:fig:DQDwField}
\end{figure}

In a perpendicular magnetic field without SOC the group of the Hamiltonian becomes the Abelian point group $\text{C}_{\text{2}}$. The only remaining symmetry operator is the total inversion $I_{xy}$ and the one-dimensional irreducible representations have either symmetric or antisymmetric base functions. The spectrum of a DQD in the perpendicular magnetic field is plotted in Fig.~\ref{ElStwB:fig:DQDwField}. The Zeeman interaction lifts the spin degeneracy and the ground state, denoted as $\Gamma_{\text{S}}^{\downarrow}$, is spin-polarized. Up to a certain magnitude of $B_z$ (about $1.5\,\text{T}$ for $2d/l_{0}=2.5$; see Fig.~\ref{ElStwB:fig:DQDwField}), the first excited state is $\Gamma_{\text{S}}^{\uparrow}$, and the spin relaxation is the transition between these two wave functions with the same orbital parts and opposite spins. For larger magnetic fields the Zeeman splitting exceeds the orbital excitation energy and the first excited state is $\Gamma_{\text{A}}^{\downarrow}$, which has the same spin polarization as the ground state. For even higher fields more states fall below $\Gamma_{\text{S}}^{\uparrow}$, which all contribute to the spin relaxation. Note that the level spacings of interest at moderate magnetic fields are smaller than the assumed valley splitting which again justifies the single valley approximation.

Within the LCSDO the single dot wave functions acquire a phase when shifted. The building blocks in Eq.~\eqref{ElSt:LCSDO} now read $\Psi_{n,l}(\mathbf{r}\pm\mathbf{d}) \,\text{exp}\!\left[\pm\text{i}eB_z \mathbf{r} \cdot \left(\hat z \times \mathbf{d} \right)/2\hbar\right]$ and we can repeat the computation of the tunneling energy with the result plotted in Fig.~\ref{ElSt:fig:tunnelingenergy}. One can see that the perpendicular magnetic field reduces the tunneling energy. This can be understood qualitatively as the renormalization of the confinement length, which is replaced by the effective (magnetoelectric) confinement length $l_{B}$, where $l_{B}^{-4} = l_{0}^{-4} + l_{\xi}^{-4}$ with the auxiliary quantity $l_{\xi} = \left(2\hbar/B_z e\right)^{1/2}$. The tunneling energy simplifies in the limit of large interdot distances to
\begin{equation}\label{ElSt:LCSDO-T-approxwField}
T \approx E_0 \dfrac{d\,l_{B}}{\sqrt{\pi}\,l_{0}^2} \,\text{e}^{-d^2 \left( 2 l_{B}^{-2} - l_{B}^2 l_{0}^{-4}\right)}.
\end{equation}
Note that for $B_z = 0$ we have $l_{B} = l_{0}$ and we obtain the results of Sect.~\ref{Sect:ElSt_Bzero}.

\section{Spin Relaxation}
\subsection{Single Quantum Dot}\label{Sect:Relax-SD}
Before we proceed to DQDs, we first discuss a single quantum dot in an in-plane magnetic field, $\mathbf{B} = B_{\parallel} \left( \cos \gamma, \sin \gamma, 0 \right)$, which already features anisotropies and relaxation rate spikes due to spin hot spots as we will see. Removing the linear SOC terms in Eq.~\eqref{Model:Hamiltonian} by a unitary transformation,\cite{PhysRevLett.87.256801, Tokatly20101104} the relaxation proceeds due to a spin-orbit-induced effective magnetic field,\cite{PhysRevLett.96.186602,PhysRevB.74.045320}
\begin{equation}\label{Relax:Beff}
B_{z}^{\text{eff}}\!=\!-B_{\parallel}\!\left[ x\!\left( \dfrac{\cos \gamma}{l_{\text{br}}} - \dfrac{\sin \gamma}{l_{\text{d}}} \right) + y\!\left( \dfrac{\sin \gamma}{l_{\text{br}}} - \dfrac{\cos \gamma}{l_{\text{d}}} \right) \right] ,
\end{equation}
which is perpendicular to the external magnetic field. The matrix element $M_{\uparrow \downarrow}$ in Eq.~\eqref{Model:rate} is proportional to this effective magnetic field, which results in the spin relaxation rate being proportional to the squared and inverse effective spin-orbit coupling length $L$,\cite{PhysRevLett.93.016601, PhysRevLett.96.186602, PhysRevB.74.045320}
\begin{equation}\label{Relax:L}
L^{-2}=l_{\text{br}}^{-2}+l_{\text{d}}^{-2}-2\sin\!\left(2\gamma\right) l_{\text{br}}^{-1} l_{\text{d}}^{-1} \, ,
\end{equation}
which is anisotropic since it depends on $\gamma$. However, the anisotropy disappears if one of the SOC lengths is dominant, particularly for $\beta=0$. Thus, experimental verification of anisotropic single dot spin relaxation would verify the existence of the generalized Dresselhaus term, Eq.~\eqref{Model:HDR}. The anisotropy is strongest if $l_{\text{br}}=l_{\text{d}}$, with the maximal rate at $\gamma=135^\circ$ and the minimal rate at $\gamma=45^\circ$.

\begin{figure}
\centerline{\psfig{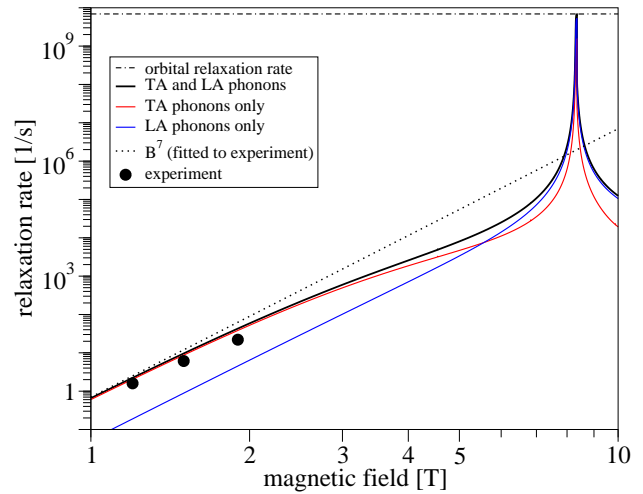}}
\caption{(color online) Spin relaxation rates of a single electron in a single quantum dot vs.~in-plane field for $\gamma=135^\circ$. The total rate (solid black line) and its contributions of the TA phonons (solid red line) and the LA phonons (solid blue line) are shown. The spin hot spot at $B_{\parallel} \approx 8.3\,\text{T}$ causes the spin relaxation rate to increase up to the orbital relaxation rate (dash-dotted line). The three dots give the experimental data of Ref.~\onlinecite{2009arXiv0908.0173H} fitted by a $B^7$ curve (dotted line).}
\label{Relax:fig:SD-relax}
\end{figure}

Figure \ref{Relax:fig:SD-relax} displays the numerical results for the spin and orbital relaxation rates with $\gamma=135^\circ$. The TA and LA phonon contributions to the total spin relaxation rate (solid black line) are given to clarify their relative importance. We find that the rate for magnetic fields up to about $5\,\text{T}$ essentially results from the TA phonons. They do not depend on $\Xi_{\text{d}}$ since the scalar product in $D_{\mathbf{Q}}^{\lambda}$, Eq.~\eqref{Model:D}, vanishes. An important observation is the strong enhancement of the total spin relaxation rate at $B_{\parallel} \approx 8.3\,\text{T}$. This is due to a spin hot spot which appears at a point at which the Zeeman splitting is equal to the level spacing of the Fock-Darwin states. The anticrossing induces a strong mixing of the spin states which abets spin relaxation. The spikes appear with equal height for any in-plane field orientation $\gamma$ as here the rate is given by the orbital relaxation rate\cite{PhysRevB.74.045320} (dash-dotted line), which is independent of $\gamma$.

In Fig.~\ref{Relax:fig:SD-relax} we also plot the spin relaxation rate as measured in Ref.~\onlinecite{2009arXiv0908.0173H}. First, the observed power dependence corresponds to the coaction of spin-orbit interactions and deformation phonons.\cite{epub1823,epub7807} The energy conservation forbids a direct electron-nuclear spin flip-flop in finite magnetic fields. This process becomes allowed if accompanied by the emission of a phonon, yielding a relaxation rate proportional to $B^{5}$.\cite{PhysRevB.64.195306} Second, the order of magnitude agreement indicates that our choice of the spin-orbit strength is realistic, even though a direct fitting is not possible (the angle $\gamma$ was not reported and the dot was not in a single electron regime).

We now derive analytic formulas for the spin relaxation rate valid for weak in-plane magnetic fields. Treating the spin-orbit coupling perturbatively, we are able to evaluate Eq.~\eqref{Model:rate} analytically. The total rate, $\Gamma_{\text{spin}} = \Gamma_{\text{spin}}^{\text{TA}} + \Gamma_{\text{spin}}^{\text{LA}}$, is given by the contributions ($\lambda' = \text{TA, LA}$)
\begin{equation}\label{Relax:RateGeneral}
\Gamma_{\text{spin}}^{\lambda'} = \mathcal{D}_{\lambda'}^2 \dfrac{m^2 l_{0}^8}{24 \pi \rho c_{\lambda'}^7 \hbar^{10}} L^{-2} (g\mu_{\text{B}}B_{\parallel})^7 .
\end{equation}
The energy parameter $\mathcal{D}_{\lambda'}^2$ reads
\begin{equation}\label{Relax:TransverseRate}
\mathcal{D}_{\text{TA}}^2 = \frac{4}{35} \Xi_{\text{u}}^2 ,
\end{equation}
and
\begin{equation}\label{Relax:LongitudinalRate}
\mathcal{D}_{\text{LA}}^2 = \Xi_{\text{d}}^2 + \frac{2}{5}\Xi_{\text{d}}\Xi_{\text{u}} + \frac{3}{35}\Xi_{\text{u}}^2 ,
\end{equation}
for the transverse and longitudinal branches, respectively. The weak versus strong magnetic field limit is determined by the conditions $\mathcal{E}_{\lambda} \ll 1$ and $\mathcal{E}_{\lambda} \gg 1$ respectively, where $\mathcal{E}_{\lambda}=g\mu_{\text{B}}B l_{B}/(\hbar c_{\lambda})$.\cite{PhysRevB.74.045320} Here, the crossover $\mathcal{E}_{\lambda}=1$ is found at $B_{\parallel}=1.4\,\text{T}$ for transverse, and $B_{\parallel}=2.6\,\text{T}$ for longitudinal acoustic phonons. Comparing with the exact numerics, we find that the error of the value of Eq.~\eqref{Relax:RateGeneral} is less than $10 \%$ up to $B_{\parallel}=0.8\,\text{T}$ for TA, and up to $B_{\parallel}=2\,\text{T}$ for LA phonons. In any case, the error is less than $5 \%$ if $B_{\parallel}<0.5\,\text{T}$.

The integral in Eq.~\eqref{Model:rate} can be done analytically only exceptionally, such as in the single dot case. Therefore, one often employs isotropically averaged deformation potentials to simplify the treatment.\cite{pop:4998,PhysRevB.48.1512,PhysRevB.81.235326} This amounts to average $D_{\mathbf{Q}}^{\lambda}$, Eq.~\eqref{Model:D}, over phonon directions distributed uniformly in three dimensions,
\begin{equation}\label{Relax:AveragingD}
\left|D_{\mathbf{Q}}^{\lambda}\right|^2 \to \left<D_{\mathbf{Q}}^{\lambda}\right>^2 \equiv \frac{1}{4\pi} \int |D_{\mathbf{Q}}|^2 {\rm d} \Omega .
\end{equation}
Here it leads to Eq.~\eqref{Relax:RateGeneral} with
\begin{equation}\label{Relax:D-TA-averaged}
\mathcal{D}_{\text{TA,iso}}^2 = 2\left<D_{\mathbf{Q}}^{\text{TA}}\right>^2 = \frac{4}{15}\Xi_{\text{u}}^2
\end{equation}
for the transverse, and
\begin{equation}\label{Relax:D-LA-averaged}
\mathcal{D}_{\text{LA,iso}}^2 =\left<D_{\mathbf{Q}}^{\text{LA}}\right>^2 = \Xi_{\text{d}}^2+\frac{2}{3}\Xi_{\text{d}}\Xi_{\text{u}}+\frac{1}{5}\Xi_{\text{u}}^2
\end{equation}
for the longitudinal contribution to the total rate. For our choice of parameters, we get $\left<D_{\mathbf{Q}}^{\text{TA}}\right>=3.29\,\text{eV}$ and $\left<D_{\mathbf{Q}}^{\text{LA}}\right>=8.44\,\text{eV}$. Comparing Eqs.~\eqref{Relax:D-TA-averaged} and \eqref{Relax:D-LA-averaged} with Eqs.~\eqref{Relax:TransverseRate} and \eqref{Relax:LongitudinalRate}, we find that the averaging, Eq.~\eqref{Relax:AveragingD}, leads to rates which are $2.3$ (TA) and, for our parameters, $1.4$ (LA) times larger than the actual rates. Note that if we use instead of Eq.~\eqref{Relax:AveragingD} the averaged deformation potentials as reported in Ref.~\onlinecite{pop:4998}, we obtain relaxation rates that differ in the low-$B$-field limit from Eqs.~\eqref{Relax:TransverseRate} and \eqref{Relax:LongitudinalRate} by a factor of $3.4$ for the TA, and $3.2$ for the LA contribution.
\begin{figure}
\centerline{\psfig{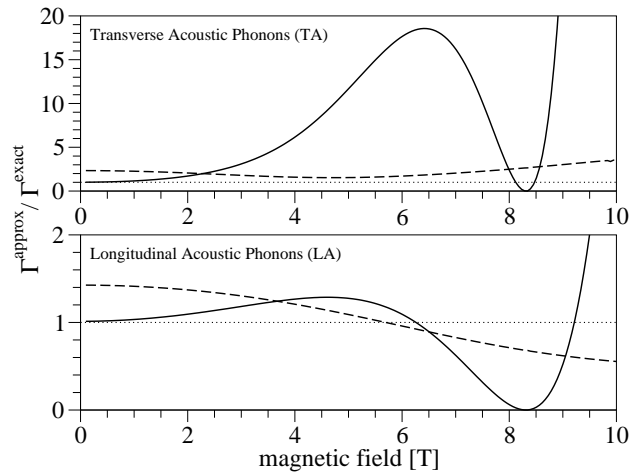}}
\caption{Ratio of the relaxation rate of the approximations and the exact numerics vs.~in-plane magnetic field. The low $B$-field limit, represented by Eqs.~\eqref{Relax:TransverseRate} and \eqref{Relax:LongitudinalRate}, is shown by the solid lines, and the numerically evaluated isotropic average approximation, Eq.~\eqref{Relax:AveragingD}, by the dashed lines. The contributions of the TA and LA phonons are given in the top and bottom panel, respectively. The magnetic field is in-plane and $\gamma=135^\circ$. The constant line at 1 (dotted) is a guide to the eye.}
\label{Relax:fig:SD-relax2}
\end{figure}

The isotropic average approximation becomes exact if the matrix element $\left| M_{\uparrow \downarrow} \right|$ is independent of the phonon direction. However, this directional invariance is not fulfilled in lateral dots which are strongly anisotropic in the perpendicular versus the in-plane direction. To assess the quality of the approximations, we compare the corresponding relaxation rates with the exact numerical result in Fig.~\ref{Relax:fig:SD-relax2} for magnetic fields up to $10\,\text{T}$. Our measure is the ratio between the rate of the approximation and of the numerics, which we plot for the TA and LA contributions separately. The parameters in Fig.~\ref{Relax:fig:SD-relax2} are identical to Fig.~\ref{Relax:fig:SD-relax}. We find that the analytical results (solid lines) deviate significantly from numerics for fields beyond the low $B$-field limit. The especially large discrepancy at around $8.3\,\text{T}$, where the ratio is close to zero, stems from the fact that the analytic approximations assume no level crossings of the initial state. Thus, it accounts for neither spin hot spots, nor the transition into excited states. Equations \eqref{Relax:D-TA-averaged} and \eqref{Relax:D-LA-averaged} result in curves parallel to Eqs.~\eqref{Relax:TransverseRate} and \eqref{Relax:LongitudinalRate}, but shifted by the discrepancy factors $2.3$ and $1.4$ for the TA and LA contributions (not shown). Numerical evaluation of the spin relaxation rates via Eq.~\eqref{Model:rate} using the average of Eq.~\eqref{Relax:AveragingD} leads to a discrepancy represented by the dashed line. We find that even in highly anisotropic (2D) lateral dots, the discrepancy factor is only of the order of 1. It is therefore expected to be legitimate to use the isotropic averaging also for more complicated dot geometries, such as the double dot, or a biased dot, where it can lead to significant simplifications.

\begin{figure}
\centerline{\psfig{file=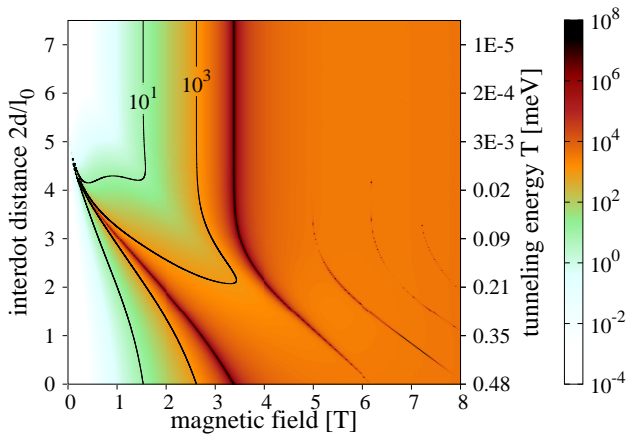,width=0.95\linewidth}}
\caption{(color online) Calculated spin relaxation rate in a DQD with tunable interdot distance in a perpendicular magnetic field. The rate is given in inverse seconds by the color with the scale on the right. The $y$-axis is calibrated in interdot distance (left) and tunneling energy at $B=0$ (right).}
\label{Relax:fig:relax-perp}
\end{figure}

\subsection{Double Quantum Dot}\label{Sect:Relax-DD}
We now move to a DQD case, where we take $2d$ as a variable parameter, noting that it could stand for either the actual separation between two dots, or a gate-tunable coupling between dots of fixed distance. From the experimental point of view, it is more convenient to characterize a double quantum dot via the tunneling energy. We plot our results with respect to the interdot distance and give also the corresponding tunneling energy at zero magnetic field computed numerically (Fig.~\ref{ElSt:fig:tunnelingenergy}, dotted line).

The spin relaxation rate in a DQD as a function of both the interdot distance in units of $l_{0}$ and the perpendicular magnetic field is shown in Fig.~\ref{Relax:fig:relax-perp}. We find that the plotted area is dominated by the spikes which come from spin hot spots and that there are no easy passages, that is there is no possibility for a fixed magnetic field to change the interdot distance from zero to infinite without passing through any of these peaks. For small fields, here $B_z < 3\,\text{T}$, we have only one relevant spike, which comes from the anticrossing of $\Gamma_{\text{S}}^{\uparrow}$ and $\Gamma_{\text{A}}^{\downarrow}$ (see Fig.~\ref{ElStwB:fig:DQDwField}). For larger magnetic fields, crossings with higher orbital states occur which may but need not lead to spin hot spots, depending on the symmetry of the crossing states.

\begin{figure}
\centerline{\psfig{file=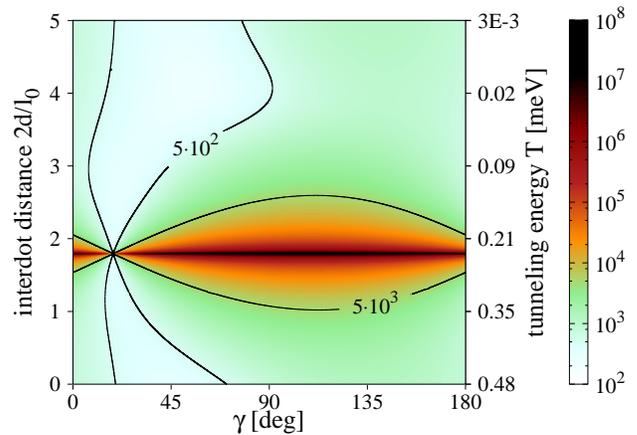,width=0.95\linewidth}}
\caption{(color online) Calculated spin relaxation rate in a DQD as a function of the interdot distance and the orientation of the in-plane magnetic field ($B_{\parallel} = 4\,\text{T}$). The rate is given in inverse seconds by color with the scale on the right. The $y$-axis is calibrated in interdot distance (left) and tunneling energy at $B=0$ (right). The dots main axis is along $\left[100\right]$.}
\label{Relax:fig:relax-ip-100}
\end{figure}

If the field is applied in the plane, the spin relaxation depends on the orientation of the magnetic field with respect to the crystallographic axes because of the interplay between Bychkov-Rashba and generalized Dresselhaus SOC. Once we rotate the coordinate system by $\delta$ around $\hat z$, the effective Zeeman field, Eq.~\eqref{Relax:Beff}, reads\cite{PhysRevLett.96.186602,PhysRevB.74.045320} {\setlength\arraycolsep{0.1em}
\begin{eqnarray}\label{Relax:Beff_rot}
\tilde B_{z}^{\text{eff}}=&-&B_{\parallel} \tilde x \left[ l_{\text{br}}^{-1} \cos(\gamma-\delta)-l_{\text{d}}^{-1}\sin(\gamma+\delta)\right]+ \nonumber \\
&+&B_{\parallel} \tilde y \left[ l_{\text{br}}^{-1} \sin(\gamma-\delta)-l_{\text{d}}^{-1}\cos(\gamma+\delta)\right] ~,
\end{eqnarray}
}where the tilted axes are such that $\tilde x$ is parallel to $\mathbf{d}$. Since the first excited orbital state $\Gamma_{\text{A}}$ transforms like $\tilde x$, only the first term in Eq.~\eqref{Relax:Beff_rot} can lead to spin hot spots for moderate magnetic fields in the intermediate regime. As an example, we plot in Fig.~\ref{Relax:fig:relax-ip-100} the spin relaxation rate in a DQD aligned along $\left[100\right]$ ($\delta=0^\circ$) in a magnetic field $B_{\parallel} = 4\,\text{T}$ varying the orientation $\gamma$ and the interdot distance. A sharp peak occurs at $2d/l_{0}=1.8$ but is intermittent at $\gamma_{\text{e}} = 18^\circ$ where the first term of Eq.~\eqref{Relax:Beff_rot} vanishes. Note that this angle, defining the easy passage, depends on the SOC lengths and is thus sample and setup dependent. An experimental determination of the easy passage angle would provide information about the relative SOC strengths via the relation $\tan\gamma_{\text{e}} = \alpha / \beta$.

\begin{figure}
\centerline{\psfig{file=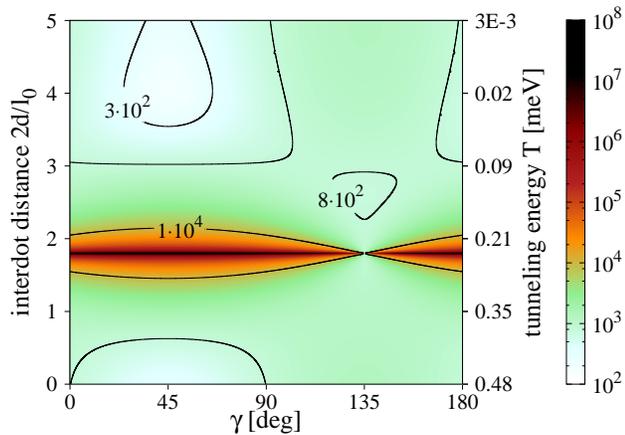,width=0.95\linewidth}}
\caption{(color online) Calculated spin relaxation rate in a DQD as a function of the interdot distance and the orientation of the in-plane magnetic field ($B_{\parallel} = 4\,\text{T}$). The rate is given in inverse seconds by color with the scale on the right. The $y$-axis is calibrated in interdot distance (left) and tunneling energy at $B=0$ (right). The dots main axis is along $\left[110\right]$.}
\label{Relax:fig:relax-ip-110}
\end{figure}

Equation \eqref{Relax:Beff_rot} shows that the easy passage depends also on the DQD orientation with respect to the crystallographic axes. We note that for a DQD with the main axis along the $\left[110\right]$ direction ($\delta=45^\circ$), the corresponding field orientation of the easy passage is universal ($\gamma_{\text{e}} = 135^\circ$), independent of the SOC strengths. We plot the spin relaxation rates for this case in Fig.~\ref{Relax:fig:relax-ip-110} for completeness.

\section{Conclusions}
We investigated the spin relaxation of a single electron confined in a quantum dot in a laterally gated Si/SiGe heterostructure. We considered the spin relaxation to be an inelastic transition in which the spin-flip is enabled by the presence of the spin-orbit interactions, while the energy difference between the initial and final state, arising from the applied magnetic field, is taken away by an acoustic phonon. We studied relaxation rates varying the interdot coupling from strong (a single dot regime) to negligible (a double dot regime), as well as varying the strength and orientation (in-plane or perpendicular) of the magnetic field. 

We adopted the single valley, effective mass, and two dimensional approximations, within which our results are numerically exact. Whereas the latter two are known to be well justified for lateral quantum dots, the single valley approximation breaks down once the valley splitting drops below the orbital energy scale, $\sim$meV, and additional states appear in the lowest part of the dot spectrum. Concerning the spin relaxation, however, these states are irrelevant as the matrix elements for the phonon induced intervalley transitions are greatly suppressed. This is so because long wavelength phonons, which arise due to the small transition energy ($q \lesssim 0.1$ nm$^{-1}$ at 3 T), are ineffective in coupling states with disparate Bloch wavefunctions ($k_v\approx 10$ nm$^{-1}$); see Eq.~(2) in Ref.~\onlinecite{PhysRevB.82.155312} and the discussion therein.

We found that the spin relaxation in Si dots is roughly comparable to that in GaAs, although it bears certain differences. Namely, in the single dot the relaxation rate in Si is proportional to $B^7$, being due to the deformation phonon potential, in contrast to the $B^5$ dependence in piezoelectric GaAs. We compared our theory with experimental data, which confirm the magnetic field power dependence and show that the spin-orbit strengths of the order of $0.1 \,\text{meV\AA}$ are to be expected in Si/SiGe quantum dots. We also derived an analytical expression for the relaxation rate treating the spin-orbit interactions in the lowest order. We find it an excellent approximation to the numerics up to magnetic fields of 1-2 T. A further simplification, the isotropic averaging, makes the analytical result to differ from the exact one by a factor of the order of 1. 

We showed that in the double dot the relaxation rate is a much more complicated function of the magnetic field and the interdot coupling, the two parameters most directly controllable experimentally. This is due to the fact that the rate is strongly influenced by spin hot spots, which occur at much lower magnetic fields in the double dot compared to the single dot. The anisotropy of the spin-orbit interactions leads to the rates dependent on the magnetic field direction with respect to the crystallographic axes. In a double dot, where the rotational symmetry of the potential is broken but the reflection symmetry is preserved, this anisotropy results in the appearance of easy passages---special directions of the external magnetic field which assure a strong suppression of the relaxation rate. From these directions the ratio of the spin-orbit strengths can be found. Compared to GaAs, in Si the easy passage position relates directly to the linear spin-orbit strengths without being influenced by the spin-orbit interaction cubic-in-momenta.

Finally, we observed that compared to GaAs, the spin relaxation rates in Si are typically 1-2 orders of magnitude smaller, as a result of the absence of the piezoelectric phonon interaction and generally weaker spin-orbit interactions.

\acknowledgments
This work was supported by DFG under grant SPP 1285 and SFB 689,
NSF under grant DMR-0706319, ERDF OP R\&D ``QUTE'', and CE SAS QUTE.

\end{document}